# Rapid Testing, Duck Lips, and Tilted Cameras: Youth Everyday Algorithm Auditing Practices with Generative AI Filters

Lauren Vogelstein, Teachers College, Columbia University, lev2124@tc.columbia.edu
Vedya Konda, University of Pennsylvania, vedyask@upenn.edu
Deborah Fields, Utah State University, deborah.fields@usu.edu
Yasmin Kafai, University of Pennsylvania, kafai@upenn.edu
Luis Morales-Navarro, University of Pennsylvania, luismn@upenn.edu
Danaé Metaxa, University of Pennsylvania, metaxa@seas.upenn.edu

**Abstract:** Today's youth have extensive experience interacting with artificial intelligence and machine learning applications on popular social media platforms, putting youth in a unique position to examine, evaluate, and even challenge these applications. Algorithm auditing is a promising candidate for connecting youth's everyday practices in using AI applications with more formal scientific literacies (i.e., syncretic designs). In this paper, we analyze high school youth participants' everyday algorithm auditing practices when interacting with generative AI filters on TikTok, revealing thorough and extensive examinations, with youth rapidly testing filters with sophisticated camera variations and facial manipulations to identify filter limitations. In the discussion, we address how these findings can provide a foundation for developing designs that bring together everyday and more formal algorithm auditing.

## Introduction and background

Artificial intelligence and machine learning (AI/ML) technologies are deeply embedded in our daily lives, influencing how we interact with and interpret the world. Various institutions have emphasized the importance of studying AI/ML to identify the necessary skills for their effective use and critique (e.g., Department of Education, 2023). While adult practices largely shape the ongoing AI/ML educational agenda, today's youth are uniquely positioned to explore everyday applications of these technologies (Anderson et al., 2023). This positions them to critically examine, evaluate, and challenge AI/ML technologies. It is essential to center youth perspectives in discussions surrounding AI/ML literacies, building on their existing skills, literacies, and creativity (Long & Magerko, 2020; Touretzky et al., 2019; Vakil & McKinney DeRoyston, 2022).

One promising approach for designing AI/ML learning environments is *algorithm auditing*, which examines algorithms from an external viewpoint. Algorithm auditing involves repeatedly querying an algorithm to observe its outputs, subsequently offering insights into its opaque mechanisms and possible impacts (Metaxa et al., 2021). Auditors can demonstrate biases within AI/ML systems by monitoring relationships between selected inputs and outputs. This is a promising avenue to support AI/ML learning because it does not require needing access to or understanding code and more closely mirrors users' experience with AI/ML technologies. Over the last decade, algorithm auditing has emerged as a vital component of research on algorithmic justice and fairness, enabling users and expert researchers alike to analyze systems for biases and harmful effects (Sandvig et al., 2014). By sharing findings publicly, they can promote social change. For example, Sweeney's (2013) audit highlighted racial bias in Google ads, revealing that searches for Black-sounding names often returned arrest record suggestions. Recently, everyday users have engaged in algorithm auditing, such as a 2020 case where a Twitter user identified a racial bias in image cropping, catalyzing a grassroots initiative that led Twitter to revise its cropping algorithm (Madland & Ofosuhene, 2022). These instances exemplify how both experts and non-experts can unearth harmful biases and advocate for meaningful technological changes (DeVos et al., 2022).

Recent studies in the United States underscore youth's capacity to critically assess AI/ML biases. Research indicates that youth leverage their identities and experiences to identify algorithmic biases (Solyst et al., 2023; Salac et al., 2023). Furthermore, studies have engaged youth in audit-like activities, introducing "evocative audits" for them to analyze computing systems' community impacts (Walker et al., 2022). Some studies adapted auditing tasks for middle school participants, asking them to evaluate Google search results and identify biases in peer-designed applications (Morales-Navarro et al., 2024). This body of work highlights the potential for adapting algorithm auditing for youth involvement. Our study adopts a microgenetic approach to explore how youth interacted with and evaluated generative AI filters.

This paper presents findings from a participatory design workshop (DiSalvo et al., 2017) where seven high school students informally investigated TikTok's generative AI filters. TikTok and similar social media platforms have gained notable popularity among youth, particularly youth of color (Anderson et al., 2023). The workshop aimed to understand how youth engage with and critique these AI/ML systems. Specifically, we analyzed a workshop activity where participants experimented with generative AI filters on TikTok—first exploring filters of their choice, then discovering the limits of these filters. We conducted a microgenetic analysis of their interactions with TikTok filters to answer two research questions: (1) How do youth approach exploring generative AI filters on TikTok? and (2) How do they comprehend the functions and limitations of these filters? We also considered the experiences and expertise they brought to this understanding. Our findings offer valuable insights into youth practices for making sense of AI/ML systems, highlighting implications for designing effective auditing learning activities.

## Methods

We conducted an eight-hour participatory design workshop over two Saturdays in the fall of 2023 with a group of seven high school youth. These participants were involved in STEM Stars, a four-year program at a science center in a Northeastern U.S. city, which brings together students from various schools for weekly STEM workshops and summer camps. Among the participants, four identified as male and three as female; six participants identified as Black, two as White, one as Latinx, and one as Asian (three participants selected multiple categories). The workshop engaged youth in four participatory inquiry and design activities (DiSalvo et al., 2017) aimed at investigating their use of AI/ML-powered applications. For this paper, we concentrate on the second activity, which consisted of 31 minutes of testing TikTok filters. Each participant received a project phone with a newly created private TikTok account to protect their privacy. They documented observations through handwritten notes and drawings while posting videos and pictures on the project account. This activity comprised two parts where participants were prompted to: (1) explore any filter(s) and draw one filter while explaining its functionality; and (2) break one of the filters and illustrate why it did not work. This study aimed to understand youth's exploratory sensemaking during play and experimentation with these filters.

Data collection involved recorded screen capture videos from the project phones used to investigate TikTok filters. This approach allowed us to see students' focuses through the phone cameras and capture the outputs of the filters (e.g., facial transformations, anime overlays). We applied micro-genetic video analysis (Derry et al., 2010; Erickson, 1982) to seven recordings from the project phones. Our focus was on the moment-by-moment interactions students had with the TikTok filters, experimenting with various filters and using their bodies as focal points for sensemaking. Through an iterative, inductive process, researchers developed a codebook grounded in youth interactions. Seven screen recordings were coded by determining the analysis scale (e.g., micro-level interactions, such as moving heads) and defining codes from three participants' recordings. We generated two primary parent codes—the name of the TikTok filter and the ways youth interacted with and tested these filters—along with various child codes. We noted "none" for instances when participants were not engaged and used "AR" to differentiate between generative AI filters and augmented reality filters that did not respond to users.

Ultimately, we created four parent codes and 24 child codes to describe youth participants' interactions with 189 filters. Three parent codes focused on interactions with filters: camera variations (angle and distance of the phone camera), facial manipulations (expression changes affecting outputs), and subject variation (altering inputs from a user to a friend's image or another drawing). Once organized, we developed color-coded data visualizations (Figure 1) to depict participant activity during this phase. The top timeline indicates TikTok filters used, while the bottom illustrates interaction types. These visualizations identified patterns in explorations, such as Danica's prolonged engagement with a single filter compared to peers (Figure 1e). We present observations on how youth tested filters' functions using various camera tools, bodily manipulations, and interactions with other objects and people. Figure 1 visualizes two students' TikTok filter explorations, highlighting key insights within limited space, annotated with selected screen captures of their experiences.

## Findings

In this section, we delve into the repertoires of practice that youth employed to examine various TikTok filters. We establish initial evidence of their iterative, extensive, and often rapid testing, highlighting their engagement with the filters. Following this, we describe three primary repertoires observed during their interactions: variations in camera use, facial manipulations, and utilizing cameras on other people and images. Each of these codes reveals how youth approached the filters—through deliberate camera positioning, playful facial manipulation, or experimentation with non-human subjects like drawings. These categories provide a framework to understand the thoughtful engagement youth displayed in their practices, which might not have been apparent under formal auditing protocols.

Collectively, the group of seven youth participants explored 189 TikTok filters in just 31 minutes, exhibiting both rapid speed and thoroughness. This suggests significant prior experience with camera phones, TikTok, and generative AI filters. Figure 1 demonstrates this rapid iterative sensemaking through visual representations of two participants' full explorations. The color shifts on the top (indicating filter changes) and the bottom (showing variations in angles or objects of focus) reflect the frequency with which youth tested different filters and multiple input variations within a single filter. For instance, Nakira utilized two main facial manipulations to swiftly explore numerous filters; she tested 14 filters in just five minutes, switching angles and facial gestures 55 times (see Figure 1a). She consistently pursed her lips and tilted her head, revealing her methodical approach to probing filters while using the same input for various outputs. However, when asked to draw one filter's functionality, she began exploring new angles and facial manipulations, indicating a broader range of exploration with that single filter (Figure 1c). Conversely, Danica also engaged in rapid testing but with a different approach. She switched between 27 filters in less than four minutes (see Figure 1e) while primarily maintaining a straight-on camera view. Her exploration was characterized by broad curiosity about multiple filters, followed by a more focused effort studying a single filter for over 10 minutes, often using a fixed pose to draw and ensure consistency (Figure 1f). Across participants, this demonstrated dynamic fluidity as they built knowledge about the filters' functions and limitations. Below, we detail the specific practices identified.

**Figure 1**
*Visualizations of Three Youth Participants' TikTok Filter Exploration*

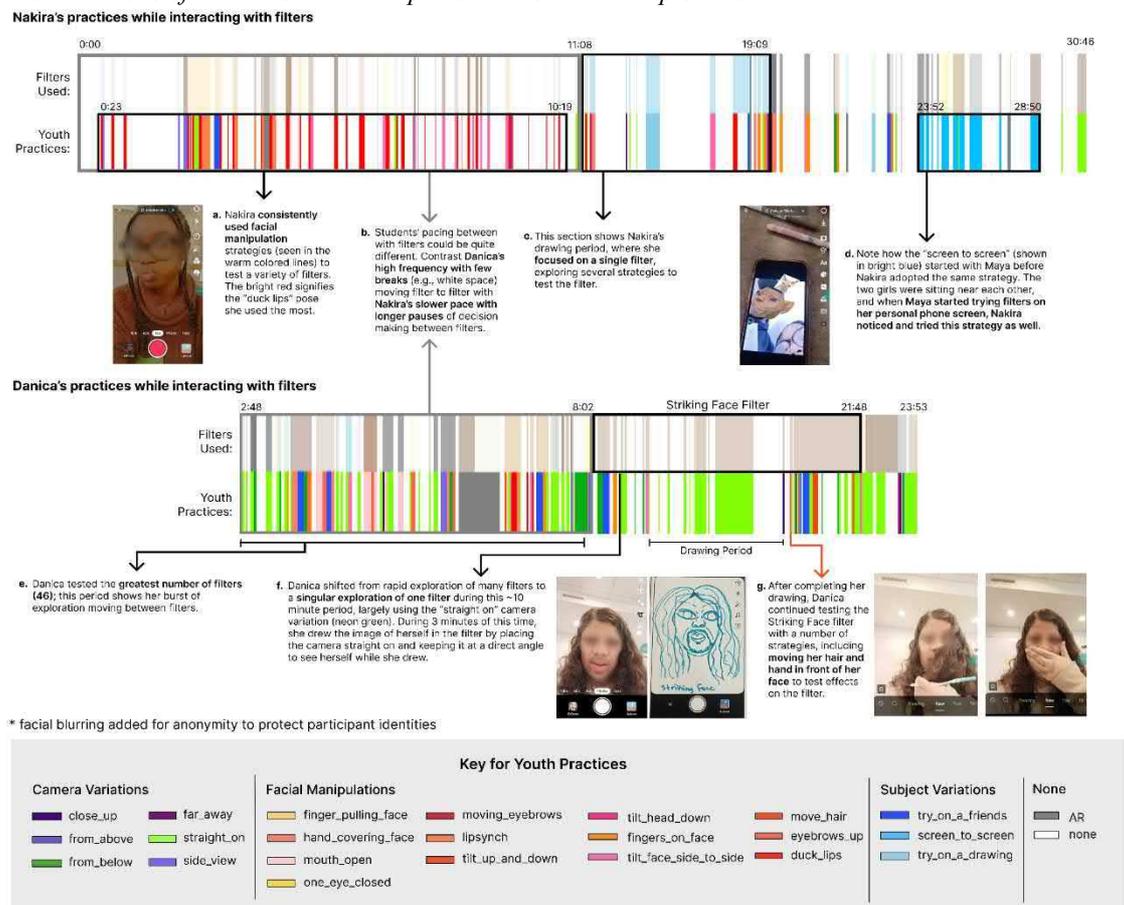

Youth employed sophisticated use of cell phone cameras, utilizing *diverse angles* and focal points to explore filters (shown in greens and purples in Figure 1). Without prior coaching, they adeptly treated the camera as a physical instrument, moving it to investigate the filters' responses from multiple perspectives. They held the camera above their heads, angled it to capture close-ups, and moved it side-to-side while keeping their faces still. This awareness allowed them to test filters with varied inputs, demonstrating extensive spatial awareness. They altered camera angles and inputs 271 times, suggesting that youth recognized how angle and framing influence filter outputs.

In addition to manipulating the camera, youth diversified their inputs by adjusting their *facial expressions*. This involved keeping the camera steady while varying gestures to examine filter behaviors. Manipulations included tilting heads, raising eyebrows, pursing lips, and other expressive gestures. Some participants further enhanced their experiments by using their hands, lip-syncing, or obscuring facial parts with hair. For example, Danica used her hair when employing the "Striking face" filter, which added facial hair effects.

By positioning her hair strategically, she altered the filter's effects in ways that could provoke discussions about gender representation (see Figure 1g). Overall, youth exploited 13 distinct facial variations a total of 213 times. While some focused on one or two specific manipulations across multiple filters, others used an array of facial variations for robust testing of individual filters.

Youth also expanded their exploration by including others in their tests, moving beyond self-experimentation to analyze filters on a broader *range of subjects*. Across all seven participants, this outward exploration occurred 88 times. In the latter part of the activity, when instructed to break a filter, many participants tested filters not only on themselves but also on two-dimensional representations: pictures from their phones (including images of people and memes) or their own drawings. For example after seeing her friends using pictures on their personal phones as inputs to test filters, Nakira engaged in this screen-to-screen strategy with several filters (see Figure 1d). This collaboration emphasizes the social and observational dimensions of learning in informal settings, where one practice sparked others' replication. Participants engaged in this multimodal practice 42 times, using drawings and images to creatively explore filter behaviors and expand their testing methods. The combination of varied camera angles, facial manipulations, and the inclusion of multiple subjects in filter analyses illustrates the thoughtful and experimental nature of youth interaction with TikTok filters. These findings highlight

## Discussion and conclusions

The focus of our research was to better understand how youth (without particular instructional directions) examined TikTok generative AI filters. Several approaches youth demonstrated, such as rapid iterative testing and different testing approaches, make the case for what we can call everyday algorithm auditing practices. We observed patterns that reflected both the participants' familiarity with similar platforms or tools and the creativity they brought to experimenting with technology. Their investigations were *extensive*, covering a tremendous number of filters (189 filters), camera variations (271 times), facial manipulations (213 times) and other people or artifacts (130 times) across just seven youth within 31 minutes. These youth were also *methodical*, if in informal ways, testing a single filter with multiple strategies or applying a couple of strategies across many filters. They *shifted their practices* based on workshop prompts to "break" a filter and draw it or in observing and incorporating peers' techniques. At the same time, they were *playful* within the ostensibly serious task of evaluating generative AI filters in TikTok, talking with each other and creatively exploring new sources of inputs by pointing their phones at other faces (in drawings, on phone pictures, on other people).

One particularly compelling finding was the extensive and rapid testing that youth conducted within a brief time period (e.g., youth sampled 20-47 filters each in 31 minutes). This rapid and iterative approach, while not identical to how experts conduct algorithm audits (Metaxa et al., 2021), shares critical features with these more professional practices. For one, the everyday auditing moved beyond one-off examinations that would be more common in casual everyday interactions on TikTok. In addition, some youth tested not just one but multiple filters using the same or similar inputs, allowing comparison of outputs. A further compelling finding was how youth creatively designed diverse inputs to examine the filters' behaviors. For instance, the youth used various facial and camera angle manipulations—approaches that have not been captured in the literature on everyday auditing. Implicit in this extensive use is their already significant expertise in using filters on TikTok, making use of cultural practices such as duck lips and various camera angles that are common in TikTok culture. The youth also used other people, objects, and spaces in the room, demonstrating thoroughness and creativity in their approaches to exploring the filters. These approaches illustrate what has been observed in everyday user auditing (DeVos et al., 2022) where non-experts bring in expertise that experts may often lack.

Our focus on youth everyday auditing practices aligns with a broader research effort to connect informal youth auditing practices with established algorithm auditing methods that expose the workings and potential biases of opaque AI systems. We identified several parallels between these youth practices and expert auditing techniques. Firstly, the youth were intentional in generating diverse inputs by manipulating camera angles and facial expressions to evaluate filters under various conditions. This intentionality mirrors formal auditing practices, which emphasize creating systematic and thoughtful inputs to assess a system's behavior. Secondly, the youth's thoroughness in cycling through numerous inputs and filters echoes the extensive data analysis in formal algorithm auditing. Thirdly, the familiarity youth had with tools like cellphone cameras allowed them to identify how different angles and distances could affect the generative AI's function, supporting the generation of hypotheses and high-quality data collection. However, aspects of youth practices differed from formal auditing methods, suggesting avenues for future design work. Specifically, the iterative manner in which youth generated inputs through interaction differed from the linear, predetermined input generation typical in formal auditing. We see significant potential in integrating youth-based practices that emphasize iterative responsiveness to AI tools with traditional scientific frameworks present in established auditing methods. This fusion could enhance auditing practices, making them more inclusive and innovative.

## Acknowledgments

We would like to thank the youth participants for sharing their generative perspectives on AI/ML as well as the STEM Stars Coordinator for their support. The analysis and writing of this paper was supported by National Science Foundation grant #2342438 for Kafai & Morales-Navarro. Any opinions, findings, and conclusions or recommendations expressed in this paper are those of the authors and do not necessarily reflect the views of the National Science Foundation, Teachers College, Utah State University, or University of Pennsylvania.